\documentclass[prl, reprint]{revtex4-1}

\pdfoutput=1
\usepackage{amsmath, amssymb, graphicx}
\usepackage[all]{xy}
\usepackage[utf8]{inputenc}
\usepackage[english]{babel}
\usepackage{times, txfonts}
\usepackage{hyperref}

\newcommand{\vphast}{{\vphantom *}}

\begin{document}

\date{September 22, 2019}

\title{Parity breakdown, vortices, and dark soliton states in a Bose
  gas of resonantly excited polaritons}

\author{S.~S.~Gavrilov}

\affiliation{Institute of Solid State Physics RAS, 142432 Chernogolovka, Russia}
\affiliation{National Research University Higher School of Economics, 101000 Moscow, Russia}
\affiliation{A.~M.~Prokhorov General Physics Institute RAS, 119991 Moscow, Russia}

\begin{abstract}
  A new mechanism of parity breakdown in a spinor Bose gas is
  predicted; it causes a single-mode state of polaritons to be
  spontaneously divided into different polarization domains which
  annihilate each other at the interface areas.  In a polariton wire,
  such interface is a dark soliton that can run in space without
  dissipation.  In a planar cavity, quantized vortices arise in which
  phase difference of orthogonally polarized components makes one
  complete turn around the core.  Coupled vortex-antivortex pairs and
  straight filaments can form in analogy to Bose-Einstein condensates
  and superconductors.  However, the rotational symmetry is broken
  even for individual vortices, which makes them interact on a large
  scale and form internally ordered structures.  These states come
  into being under resonant excitation if the spin coupling rate
  significantly exceeds the decay rate.
\end{abstract}

\maketitle

Equilibrium Bose-Einstein condensates obey an autonomous
U(1)-invariant wave equation, but the very onset of macroscopic
coherence is accompanied by symmetry breaking so that all particles
share the same spontaneously chosen phase.  When the symmetry has
broken, the underlying phase freedom reveals itself in quantized
vortices, topological excitations arising because of a weak Coulomb
repulsion of particles.  They were observed in various systems
including quantum liquids~\cite{Pitaevskii-book-2016,
  Leggett-book-06}, superconductors~\cite{Blatter94}, cold
atoms~\cite{Matthews99,Weiler08}, and
microcavities~\cite{Scheuer99,Lagoudakis08}.  Of particular interest
are cavity polaritons, mixed states of photons and excitons formed in
a thin quasi-two-dimensional layer of a semiconductor
microcavity~\cite{Yamamoto-book-2000,Kavokin-book-17}.  Their coherent
states originate in two ways, (i) via Bose-Einstein condensation from
a nonresonantly pumped excitonic reservoir to the ground state or,
thanks to the photonic component, (ii) directly under resonant and
coherent optical driving~\cite{Baas06}.  In contrast to equilibrium
systems, the directly driven condensate displays forced oscillations
and its phase is not free but imposed by the pump wave.  For this
reason all known ways to excite vortices resonantly involve explicit
patterning of the incident pump beam and/or intracavity resonance
energy~\cite{Whittaker07, Marchetti10, Krizhanovskii10, Sanvitto10,
  Sanvitto11, Dominici15, Boulier15, Dominici18, Liew07,
  Liew08-prl-vort}.  At the same time, nobody has seen vortices formed
in a homogeneous Bose system continuously driven by a plane
electromagnetic wave.  This phenomenon is the subject of the current
Letter.

We report a new kind of quantized vortices that originates
specifically under the conditions of resonant excitation owing to
spontaneous breakdown of the spin-reversal symmetry (parity).  As will
be shown below, parity breaks down when particles of opposite spin are
linearly coupled.  A homogeneous, spin-symmetric, and symmetrically
driven initial state of a Bose gas is then divided into large-scale
domains that differ in the way of symmetry breaking.  Two equally
possible steady states have opposite phases, so they annihilate each
other and, consequently, the boundary between different domains is
dilute and highly unstable.  It gives birth to vortices in a
two-dimensional system and dark solitons in a one-dimensional system.
In some respects these excitations are very similar to their
equilibrium counterparts, for instance, vortex-antivortex dipoles are
easily formed on reaching a crossover point in analogy to atomic gases
(\cite{Neely10}).  At the same time, vortices in the driven system can
interact on a scale largely exceeding their healing length and thus
form complex patterns with internal ordering.  The polarization
properties of such vortices are also uncommon and do not fit into the
conventional row of the half-~\cite{Rubo07,Lagoudakis09}, full-, and
spin-vortex states~\cite{Dominici18} as well as linear-polarization
vortices~\cite{Liew07,Liew08-prl-vort}.

Let us consider a planar polariton system excited at normal incidence
(axis $z$) by a coherent light wave.  Polaritons have two spin states
matching two opposite directions of total angular momentum
($J_z = \pm 1$); they correspond to right- and left-handed circular
polarizations of light.  Opposite spin states do not interact pairwise
if the driven mode is far below the exciton
level~\cite{Sekretenko13-10ps,Vladimirova10}.  However, they can be
linearly coupled owing to lifted degeneracy of eigenstates with
orthogonal polarizations, which takes place in the presence of a
mechanical
stress~\cite{Gavrilov13-apl,Sekretenko13-fluct,Gavrilov14-prb-j} or
constant magnetic field parallel to the cavity surface.  Within the
mean-field approach, this system is described by two amplitudes
$\psi_\pm(\mathbf{r}, t)$ obeying the nonautonomous (``driven'')
variant of the Gross-Pitaevskii equation~\cite{Kavokin-book-17},
\begin{equation}
  \label{eq:gp}
  i \hbar \frac{\partial \psi_\pm}{\partial t} =
  \left( \hat E - i \gamma
  + V \psi_\pm^* \psi_\pm^\vphast
  \right) \psi_\pm^\vphast
  + \frac{g}{2} \psi_\mp^\vphast
  + f_\pm^\vphast e^{-i E_p t / \hbar}.
\end{equation}
Here, $\hat E = \hat E(-i \hbar \nabla)$ is the energy operator
determined by the lower-polariton dispersion law $E(k)$ that is nearly
parabolic for small in-plane wave numbers $k$; $\gamma$ is the decay
rate, $V > 0$ is the polariton-polariton interaction constant, $g/2$
is the spin coupling rate.  When $\psi_\pm \to 0$ and the cubic terms
are negligible, the system is diagonalized by the unitary
transformation $\psi_\pm = (\psi_x \mp i \psi_y) / \sqrt{2}$.  The
eigenstates at $f_\pm = 0$ are orthogonally polarized and their energy
levels are $E_{x,y}(k) = E(k) \pm g/2$, so that $g = E_x - E_y$ for
each $k$.  The last term in Eq.~(\ref{eq:gp}) represents an effective
pump force, where $f_+$ and $f_-$ are proportional to the respective
polarization components of the external electric field.  The pump
frequency $E_p / \hbar$ is supposed to be moderately close to the
resonance frequency of the driven polariton state with zero~$k$.

Throughout this work we consider the case of spin-symmetric excitation
($f_+ = f_- = f$), so that the equations for $\psi_+$ and $\psi_-$ are
exactly the same.  Since the model is homogeneous, they always have
one-mode solutions of the form
$\psi_\pm (t) = \bar\psi_\pm e^{-i E_p t / \hbar}$, similar to forced
oscillations of a dissipative pendulum.  Amplitudes $\bar\psi_\pm$
obey time-independent equations
\begin{equation}
  \label{eq:steady-state}
  \left(
    D + i \gamma - V |\bar\psi_\pm|^2
  \right) \bar\psi_\pm
  - \frac{g}{2} \bar\psi_\mp = f,
\end{equation}
where $D = E_p - E(k\,{=}\,0)$ is the pump detuning from the mean
(unsplit) ground-state level.  This system is
multistable~(\cite{Gippius07}).  Clearly it has spin-symmetric
solutions with $\psi_+ = \psi_-$ for each $f$, which we henceforth
denote as the $\Pi$ states.  They are linearly polarized in the $x$
direction, i.\,e., in the same way as the incident pump.  The
orthogonal $(y)$ component is not excited, so the full spinor problem
is reduced to a scalar problem with different pump detuning
$D' = D - g/2$.  In particular, the $\Pi$ states exhibit intrinsic
bistability when $D' > \sqrt{3} \gamma$~\cite{Elesin73,Baas04-pra}.

In order to demonstrate different types of solutions, let us equate
the left sides of Eqs.~(\ref{eq:steady-state}), which gives
\begin{equation}
  \label{eq:ratio}
  \frac{\bar\psi_-}{\bar\psi_+} =
  \frac
  {D + g/2 + i \gamma - V |\bar\psi_+|^2}
  {D + g/2 + i \gamma - V |\bar\psi_-|^2}.
\end{equation}
It is seen that if $V |\bar\psi_\pm|^2 = D + g/2$, then
$\bar\psi_\mp \propto \gamma$, so that the circular-polarization
degree
$S_1 = (|\bar\psi_+|^2 - |\bar\psi_-|^2) / (|\bar\psi_+|^2 +
|\bar\psi_-|^2)$ nearly reaches $\pm 1$ at small $\gamma$.  This kind
of steady states will be denoted as $\Sigma_\pm$.  They were
experimentally observed in a microcavity with
$g \gtrsim \gamma$~\cite{Gavrilov13-apl, Gavrilov14-prb-j,
  Sekretenko13-fluct}.

The system also has another way of parity breaking which comes into
play at greater $g / \gamma$ and is directly responsible for vortices
and dark solitons.  Suppose $V |\bar\psi_\pm|^2 = D + g/2 \pm \delta$,
so that Eq.~(\ref{eq:ratio}) turns into
\begin{equation}
  \label{eq:ratio_omega}
  \frac{\bar\psi_-}{\bar\psi_+} =
  \frac
  {i \gamma - \delta}
  {i \gamma + \delta}.
\end{equation}
Using these formulae, one can simplify Eq.~(\ref{eq:steady-state}) for
$\bar\psi_+$.  After taking the absolute value of both sides, it
yields
\begin{equation}
  \label{eq:eq_for_delta}
  \left(
    \gamma^2 + \delta^2 + \frac{\gamma^2 g^2}{\gamma^2 + \delta^2}
  \right)
  \left( D + \frac{g}{2} + \delta \right)  = V f^2.
\end{equation}
If $|\delta|$ is much smaller than $g$ and $D$, one can drop $\delta$
in the second parentheses of Eq.~(\ref{eq:eq_for_delta}) which is then
reduced to a quadratic equation for $\gamma^2 + \delta^2$.  The latter
has positive roots starting with pump intensity
\begin{equation}
  \label{eq:threshold}
  f_1^2 = \frac{2 \gamma g}{V} \left( D + \frac{g}{2} \right),
\end{equation}
where $\gamma^2 + \delta^2 = \gamma g$.  The assumed smallness of
$|\delta|$ is justified when $\gamma \ll g \sim D$, in which case the
overall consideration based on Eq.~(\ref{eq:ratio_omega}) is
self-consistent.  Near the threshold point $f = f_1$ we have
$\delta / g \to 0$ and $\delta / \gamma \to \infty$ if
$\gamma / g \to 0$.  In accordance with Eq.~(\ref{eq:ratio_omega}),
$\bar\psi_- / \bar\psi_+ \to -1$, so that the field is polarized
linearly in the $y$ direction.

\begin{figure}
  \centering
  \includegraphics[width=\linewidth]{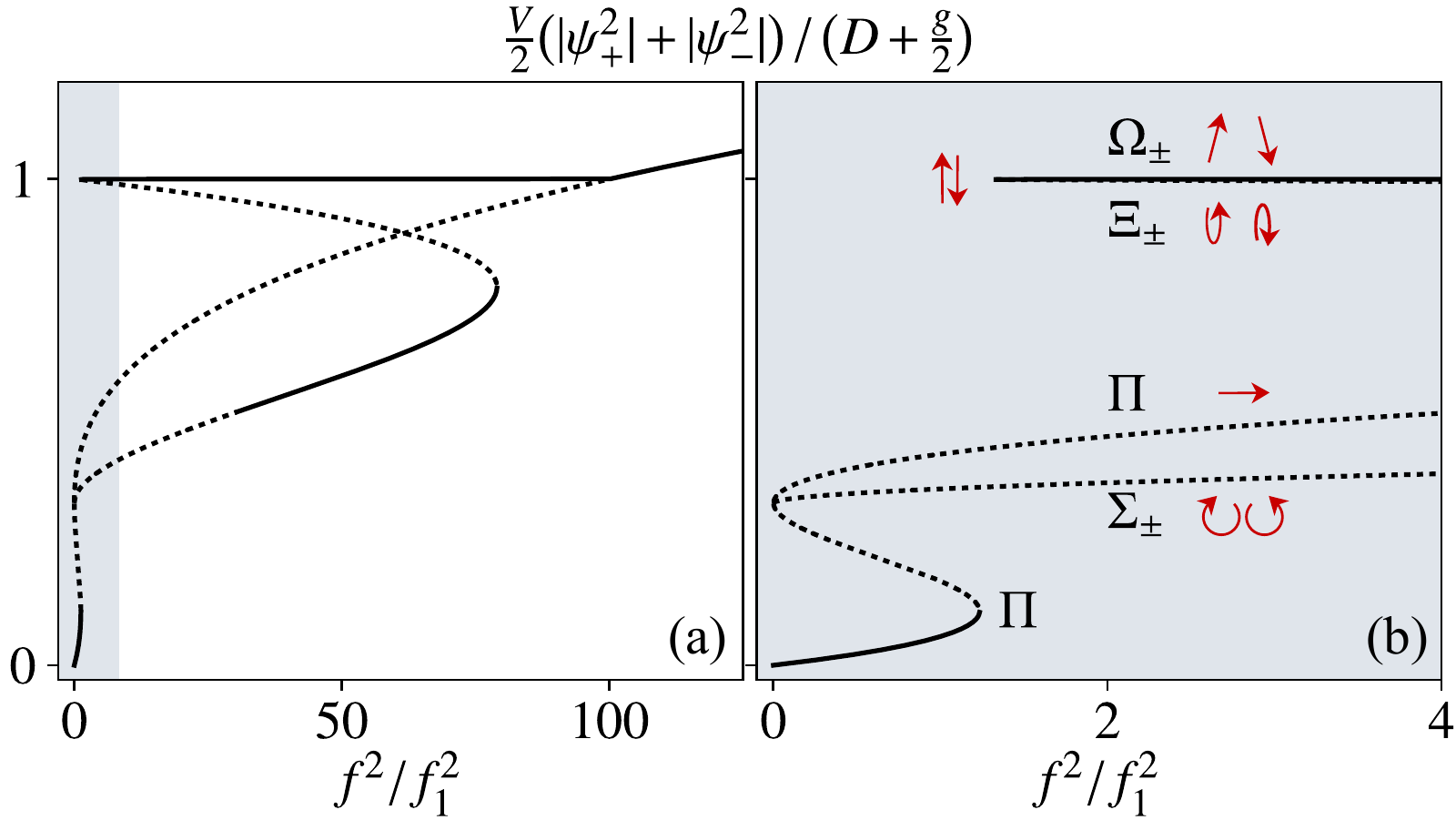}
  \caption{\label{fig:steady_states} One-mode response functions
    [solutions of Eqs.~(\ref{eq:steady-state})] at
    $\gamma = 5~\mu\mathrm{eV}$ and $g = D = 1$~meV, so that
    $f_2^2 / f_1^2 = g / 2 \gamma = 100$.  Dotted lines represent the
    solutions which are unstable even in the one-mode limit for
    $k = 0$.  Panel (b) shows a magnified interval of (a) at small
    $f$; arrows and ellipses schematically indicate polarization
    states.}
\end{figure}

Being very small at the threshold, $|\delta|$ decreases still further
with increasing $f$ for one of two pairs of solutions of
Eq.~(\ref{eq:eq_for_delta}).  At $f \gg f_1$ we have
$\delta(f) = \pm \gamma \sqrt{f_2^2 / f^2 - 1}$, where
$f_2^2 = (g / 2 \gamma) f_1^2$ is the upper threshold at which
$|\delta|$ eventually turns to zero.  Below this point, the condensate
has constant total intensity $I = |\bar\psi_+|^2 + |\bar\psi_-|^2$ but
varying polarization direction.  The $(x,y)$ polarization degree,
$S_2 = (|\bar\psi_x|^2 - |\bar\psi_y|^2) / I$, shows a linear increase
as a function of $f^2$ and ranges from about $-1$ at $f = f_1$ to $+1$
at $f = f_2$.  The solutions with mutually opposite $\delta$ are
distinguished by the sign of the ``diagonal'' linear polarization
$S_3 = (\bar\psi_x^* \bar\psi_y^\vphast + \bar\psi_y^*
\bar\psi_x^\vphast) / I$.  Since the length of the Stokes vector
$(S_1, S_2, S_3)$ is unity and its circular-polarization component
$S_1 = \delta / (D + g/2)$ is negligible, we have
$S_3 \approx \pm \sqrt{1 - S_2^2}$ and, in particular,
$S_3 \approx \pm 1$ at $f^2 = (f_1^2 + f_2^2) / 2$.  Let us denote
this doublet of solutions as $\Omega_\pm$.  At the upper threshold,
the $\Omega_\pm$ branches degenerate into the singlet $\Pi$ branch, as
it is seen from Fig.~\ref{fig:steady_states} in which all solutions of
Eq.~(\ref{eq:steady-state}) are represented.

In a cousin doublet of solutions, henceforth $\Xi_\pm$, the value of
$|\delta|$ grows with $f$, so the field gradually acquires noticeable
right- or left-handed circular polarization.  Since then
ratio~(\ref{eq:ratio_omega}) does not satisfy
Eqs.~(\ref{eq:steady-state}) even approximately.  As seen in
Fig.~\ref{fig:steady_states}, the intensity of the $\Xi$ states
decreases with increasing $f$, which is indicative of instability.
Sooner or later the $\Xi$ doublet meets the $\Sigma$ doublet and they
both terminate.

The key point is that the $\Omega$ doublet is the only possible type
of steady states in a wide interval of $f^2$ when
$\gamma \ll g \sim D$.  The instabilities of the $\Pi$ and
$\Sigma_\pm$ states were investigated earlier, and here we only
briefly remind the main results.  The spin-symmetric $\Pi$ state is
unstable because an indefinitely small imbalance of $|\psi_+|$ and
$|\psi_-|$ makes the greater component grow further by simultaneously
suppressing the minor one~\cite{Gavrilov13-apl, Gavrilov14-prb-j,
  Sekretenko13-fluct, Gavrilov14-jetpl-en, Gavrilov16-helix}.  This
occurs at $g \gtrsim \gamma$ and leads to one of the $\Sigma_\pm$
states.  After one of the spin components has been suppressed by the
other, a significant increase in the pump intensity is required for
driving it up again, so that the length of the $\Sigma$ branches in a
diagram like Fig.~\ref{fig:steady_states} is quite great.  Notice as
well that the Josephson oscillations (\cite{Shelykh08-j}) are not
possible in our system so long as both spin components have the same
``forced'' frequency $E_p / \hbar$ and, consequently, their phase
difference does not vary with time.  On the other hand, the pair
interaction conserves spin and, taken alone, it also cannot help
restore spin symmetry.  However, the $\Sigma$ branches lose stability
with respect to a higher-order loop interaction process
\begin{equation}
  \label{eq:diag:loop}
  \begin{gathered}
    \xymatrix{
      {} & {} & *+[Fo]{g} & {} & {} \\
      {} &
      *+[Fo]{V}
      \ar@{~}[ul]^{\pm}
      \ar@{~}[dl]_{\pm}
      \ar@{~}[ur]_{\pm}
      \ar@{~}[dr]^{\pm}
      & {} &
      *+[Fo]{V}
      \ar@{~}[ul]^{\mp}
      \ar@{~}[dl]_{\mp}
      \ar@{~}[ur]_{\mp}
      \ar@{~}[dr]^{\mp}
      & {} \\
      {} & {} & *+[Fo]{g} & {} & {}}
  \end{gathered}
\end{equation}
---which simultaneously enables spins to be reversed and lifts the
frequency degeneracy.  Even when the condensate at $E = E_p$ has a
perfectly circular polarization, new energy levels with different
polarizations get populated at $E \approx E_p \pm D$ in a finite
interval of $k$ around $k = 0$.  This process starts at
$g \approx 4 \gamma$~\cite{Gavrilov17-jetpl-en} and leaves no one-mode
solutions within a certain range of $f^2$, which results in
turbulence~\cite{Gavrilov16}, periodic spin networks, and chimera
states~\cite{Gavrilov18-prl}.

Given that $g \sim D > 0$, a decrease in the decay rate $\gamma$
broadens the interval of $f^2$ in which all of the $\Pi$ and
$\Sigma_\pm$ states are forbidden.  Simultaneously it lowers the
critical point $f_1^2 \propto \gamma$ where the $\Omega_\pm$ states
appear.  They have the greatest intensity for each $f < f_2$ and are
always stable.

The $\Pi \to \Omega$ transition is qualitatively different from the
$\Pi \to \Sigma$ transition observed at $g \sim \gamma$.  First, the
superposition of $\Omega_+$ and $\Omega_-$ has extremely low intensity
near $f = f_1$.  To make it clear, notice that replacing $\psi_+$ with
$\psi_-$ and vice versa turns $\Omega_+$ into $\Omega_-$ and
$\Sigma_+$ into $\Sigma_-$ in view of symmetry.  Thus, the average
state for each doublet is spin-symmetric and must have the $x$
polarization direction $(S_2 = +1)$.  We have found, however, that the
$\Omega_\pm$ states are nearly $y$-polarized at $f = f_1$, so they
simply annihilate each other, which also applies to the $\Xi_\pm$
pair. That is why a turnover point between $\Omega_+$ and $\Omega_-$
can behave like a dark soliton or vortex core.

Second, notice that the outcome of the $\Pi \to \Sigma_\pm$ transition
at $g \sim \gamma$ is determined in its very beginning by the sign of
$|\psi_+| - |\psi_-|$.  By contrast, nothing determines the outcome of
the $\Pi \to \Omega_\pm$ transition near $f = f_1$ at its early stage,
because processes of the type~(\ref{eq:diag:loop}) make a continuum of
different $k$-states populated concurrently.  After the spin symmetry
has broken, the system goes through the stage of very strong
spatiotemporal disorder.  That is why vortices and dark solitons arise
truly spontaneously, i.\,e., owing to infinitesimal fluctuations
rather than finite seeding inhomogeneities.

The following numerical experiments are performed with typical
parameters of a GaAs-based microcavity.  The exciton-photon detuning
at $k = 0$ is zero, the full Rabi splitting is 10~meV, the exciton
mass is much greater than the photon mass $\epsilon E / c^2$, where
$\epsilon = 12.5$ and $E = 1.5$~eV.  Interaction constant $V$ can be
chosen arbitrarily, as it only determines
threshold~(\ref{eq:threshold}) for given $\gamma$, $g$, and $D$.  A
white-noise term is added to the right side of Eqs.~(\ref{eq:gp}) to
simulate quantum fluctuations; taken alone, it brings about a $10^9$
times smaller average population $|\psi_\pm|^2$ than that induced by
the regular pumping.  The pump is smoothly switched on during 0.1~ns
and then held constant.  The boundary conditions are arranged by means
of a sharp increase in the polariton energy (which renders the pump
off-resonant) or decay rate.

\begin{figure}
  \centering
  \includegraphics[width=\linewidth]{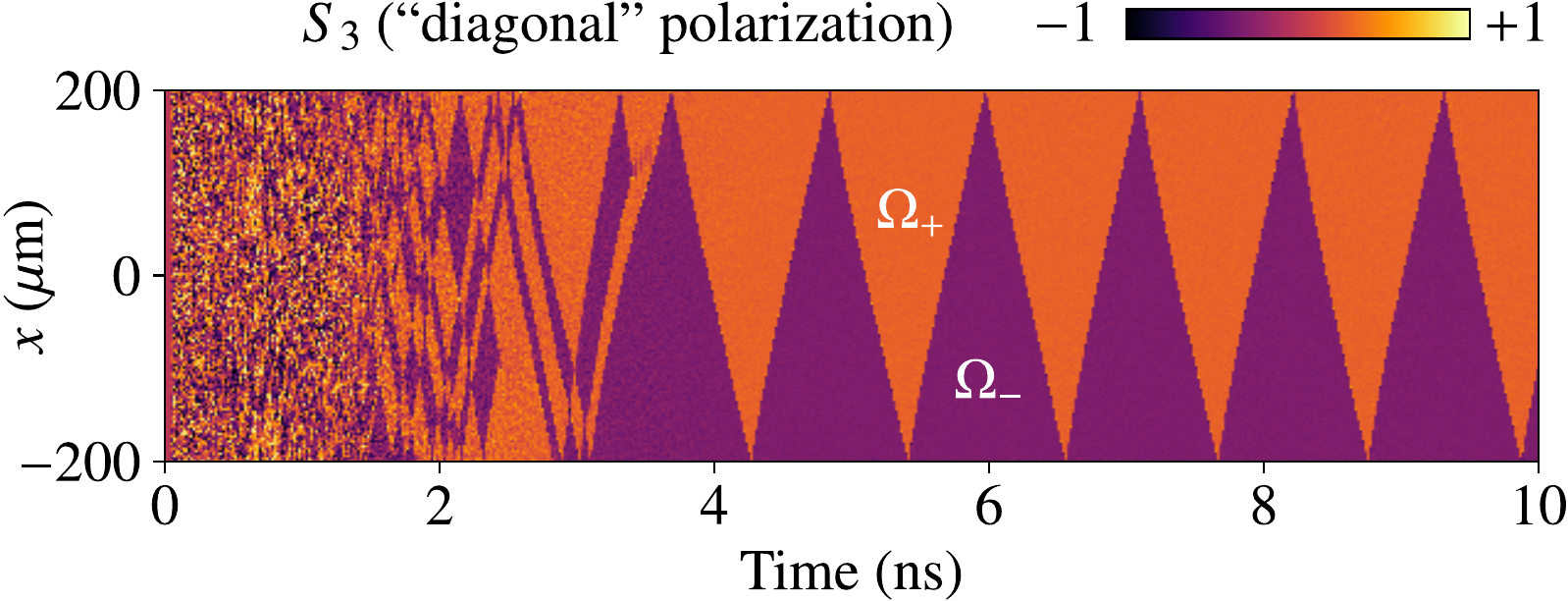}
  \caption{\label{fig:soliton} Solution of Eqs.~(\ref{eq:gp}) at
    $\gamma = 1~\mu\mathrm{eV}$, $g = D = 1$~meV, and
    $f^2 / f_1^2 = 11$ in the one-dimensional case.  The boundary
    between the $\Omega_\pm$ states behaves as a dark soliton; it runs
    with a constant speed and gets reflected from the 2-meV high
    potential walls at $x = \pm 200~\mu$m.}
\end{figure}

\begin{figure}[!b]
  \centering
  \includegraphics[width=\linewidth]{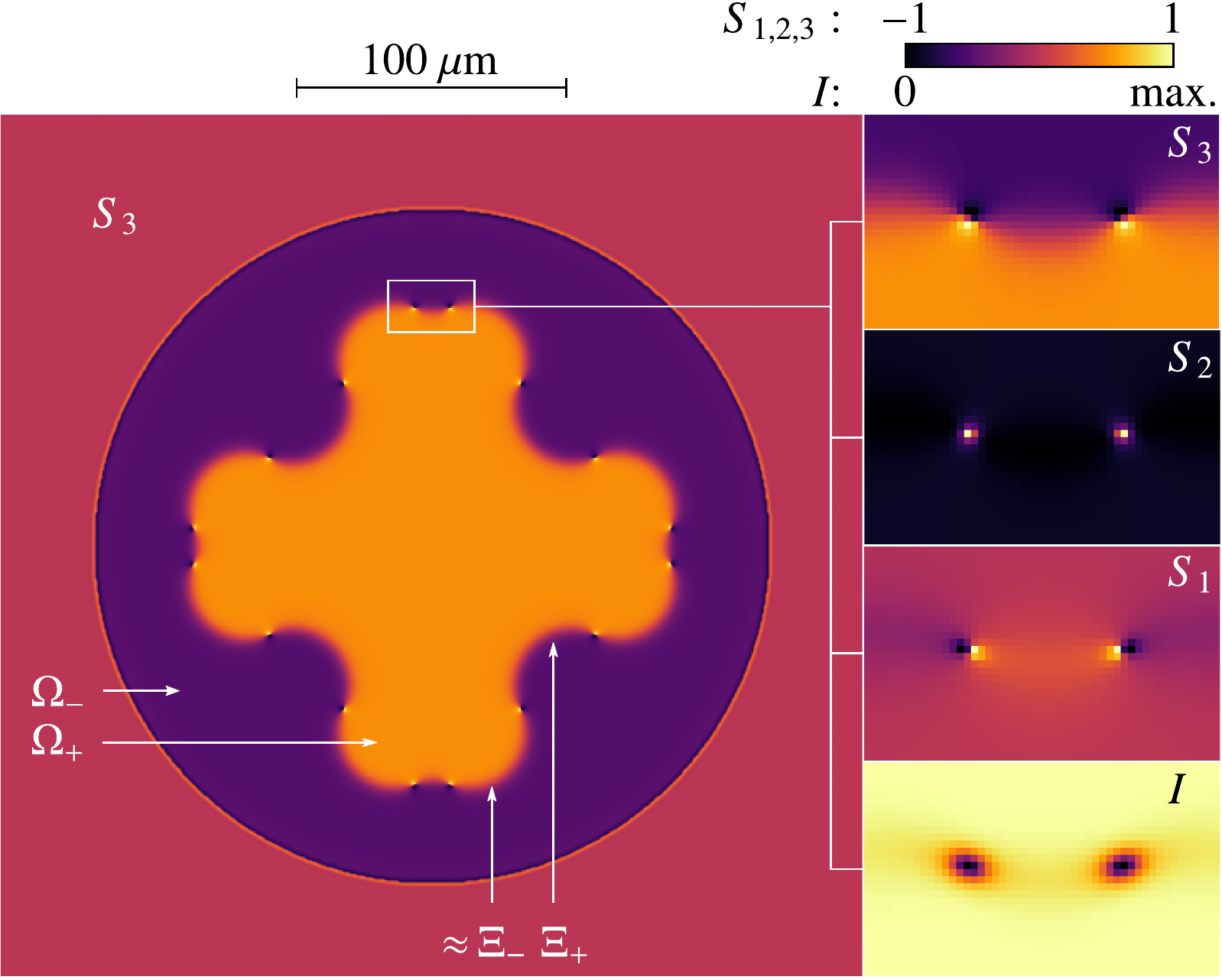}
  \caption{\label{fig:cross} Intensity $I$ and polarization degrees
    (normalized Stokes parameters) $S_{1,2,3}$ for a steady solution
    of Eqs.~(\ref{eq:gp}) formed in about 1~ns at
    $f^2 / f_1^2 \approx 6.8$.  Parameters $\gamma$, $g$, and $D$
    correspond to Fig.~\ref{fig:steady_states}.  In order to avoid a
    chaotic stage of evolution and retain symmetry, the initial
    conditions were biased to the $\Omega_+$ and $\Omega_-$ states at
    $|\mathbf{r}| \le 20~\mu$m and $|\mathbf{r}| > 20~\mu$m,
    respectively, $\mathbf{r} = 0$ being the grid center.  To ensure
    zero boundary conditions, $\gamma$ is increased up to 4~meV at
    $|\mathbf{r}| = 125~\mu$m.  Insets show a magnified fragment which
    contains two vortices with opposite topological charges.}
\end{figure}

Figure~\ref{fig:soliton} shows a $400~\mu$m long one-dimensional
polariton system with potential walls near the boundaries.  After the
stage of disorder, this system nearly approaches equilibrium but
remains two-component; it tends to be in the $\Omega$ state but does
not prefer one of the $\Omega$ twins over the other.  Each pair of the
$\Omega_\pm$ domains are separated by a dark soliton, point-like
interface at which total intensity $I$ significantly drops down.  When
$\gamma / g$ is especially small and $f$ close to $f_1$, such soliton
can endlessly run and even reflect from the potential walls.
Colliding solitons cancel each other, thus, only one can survive in
the long term, resulting in a very unusual type of spatiotemporal
self-pulsations seen in Fig.~\ref{fig:soliton}.  Increasing $f$ makes
solitons fixed in space, so that the field becomes randomly divided
into static polarization domains with alternating $\Omega_\pm$.  In
the latter case, which is also typical of comparatively large values
of $\gamma / g$, opposite $\Omega_+$ and $\Omega_-$ states can be
manually toggled back and forth by additional short-term excitation
pulses.

The two-dimensional case (Fig.~\ref{fig:cross}) is essentially more
complex.  Of particular interest is the boundary between the
$\Omega_\pm$ states, its shape and polarization.  To skip the
turbulent stage, we have used rotationally symmetric initial
conditions biased to the $\Omega_+$ and $\Omega_-$ states,
respectively, inside and outside a circle located at the grid center.
Accordingly, the pump was not ``switched on'' smoothly but had just a
fixed amplitude.  Figure~\ref{fig:cross} shows the steady state to
which such a system came after about 1~ns of evolution.

Here, the borderline between $\Omega_+$ and $\Omega_-$ comprises 16
curved segments whose polarizations appear to be very close to the
$\Xi$ states.  They alternate each other and tend to move in opposite
directions from $\Omega_\mp$ to $\Omega_\pm$ for $\Xi_\pm$.  Each
turnover point between $\Xi_+$ and $\Xi_-$ naturally carries a vortex
core and has very low intensity.  Neighboring vortices with opposite
rotation directions balance each other and help stabilize the system.
The steady states like that represented in Fig.~\ref{fig:cross} are
always internally balanced.  However, in many cases the system does
not come to a steady state even in tens of nanoseconds.

The Stokes vector components shown in Fig.~\ref{fig:cross} allow one
to deduce the phase distribution around the core.  It is seen that
$S_1 \sim \pm 1$ and $S_3 \sim \pm 1$, respectively, along the tangent
and the normal to the borderline at the core point.  The signs of
$S_3$ and $S_1$ match the $\Omega_\pm$ domains and $\Xi_\pm$ segments.
A purely circular polarization $S_1 = \pm 1$ implies that (i)
$|\psi_+| = |\psi_-|$ and (ii) the $(x,y)$ phase difference
$\Delta\phi = \arg(\psi_x^* \psi_y^\vphast)$ is equal to $\pi / 2$ or
$3 \pi / 2$, whereas $S_3 = \pm 1$ implies $\Delta\phi = 0$ or $\pi$.
Thus, $\Delta \phi$ makes an angle of $\pm 2 \pi$ around the core.  At
the same time, the orientation of the $S_{1,3}$ poles of each vortex
is certain and corresponds to the long-range symmetry breaking.
Static vortices of this kind can hardly bear topological charges other
than~$\pm 1$.

When two vortices of opposite charge come within short distances of
each other, the borderline segment they bound becomes straight.  In
fact many adjacent vortices may cancel their high-$S_1$ poles and
unite into straight $\Omega$-neutral filaments which do not move in
space.  At larger $f / f_1$, when the average of $\Omega_+$ and
$\Omega_-$ acquires a noticeably nonzero $I$, this new kind of the
borderline is highly preferred over the $\Xi$ states.  The whole
picture turns out to be steadily divided into several $\Omega_\pm$
domains by straight yet randomly directed lines which are analogous to
fixed dark solitons formed in the one-dimensional system.  The same
occurs at comparatively large $\gamma / g$ when $\Omega_+$ and
$\Omega_-$ do not annihilate each other completely even at $f = f_1$.

\begin{figure}
  \centering
  \includegraphics[width=\linewidth]{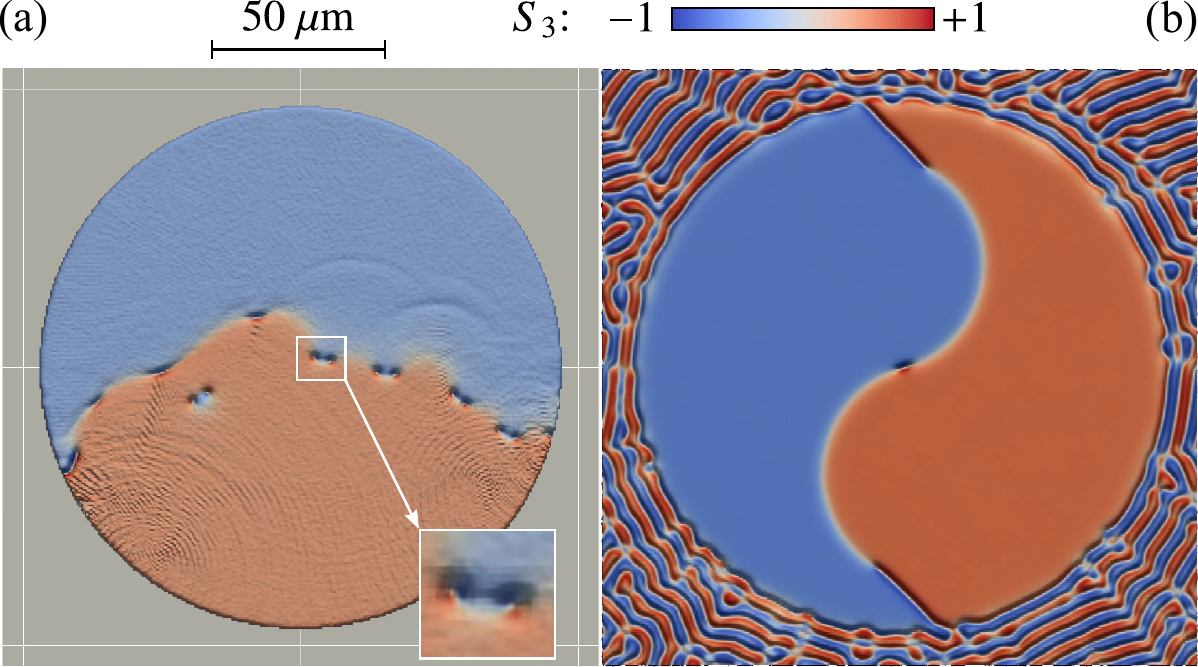}
  \caption{\label{fig:paired} Snapshots of unsteady solutions of
    Eqs.~(\ref{eq:gp}).  Parameters $\gamma$, $g$, and $D$ correspond
    to Fig.~\ref{fig:steady_states}.  The value of $f^2 / f_1^2$ is
    6.8 for (a) and 13 for (b).  The initial conditions are
    homogeneous (zero) in both cases.  In (a), the boundary conditions
    are zero thanks to a high potential wall at
    $|\mathbf{r}| = 75~\mu$m.  No walls are present in (b), but
    $\gamma$ is increased up to 0.2~meV at the same $|\mathbf{r}|$,
    which results in a filamented field distribution~\cite{Gavrilov16}
    in the outer area; the boundary conditions are periodic.  The
    obtained pattern steadily rotates clock-wise and makes a complete
    turn in 2.6~ns.  Supplemental Materials \cite{Note1, Note2}
    contain video files showing the evolution of (a) and (b).}
\end{figure}

Figure~\ref{fig:paired} and video files \footnote{Supplemental
  material showing the evolution of Fig.~\ref{fig:paired}(a),
  \url{https://drive.google.com/open?id=1R3L5zq_-glJ4VODxrT7sFwXiPnxMOcCG}}
and \footnote{Supplemental material showing the evolution of
  Fig.~\ref{fig:paired}(b),
  \url{https://drive.google.com/open?id=1LltcihsBJ_FyCxJG_bveh5ByILizI7rO}}
illustrate a crossover from the filaments to individual vortices
analogous to the transition discovered by Berezinskii, Kosterlitz, and
Thouless (BKT)~\cite{Kosterlitz73,Hadzibabic06}.  Both patterns (a)
and (b) form spontaneously; the initial conditions are zero.  In
Fig.~\ref{fig:paired}(a), the field is enclosed by a ring-shaped
potential wall, so that polaritons can reflect from the boundary
similar to Fig.~\ref{fig:soliton}.  The $\Xi$-state borderline has
$S_2 \sim -1$ and runs in space freely.  However, at some places it
occasionally turns into the said filaments with smaller $I$ and larger
$S_2$ sandwiched between the states with $S_3 \sim \pm 1$.  The
filaments slow down and break into the vortex-antivortex molecules
which have small lifetime but freely penetrate into the $\Omega$
domains, in contrast to single vortices which are merely pinned to the
borderline.

Instead of erecting a wall, in Fig.~\ref{fig:paired}(b) we have
increased the decay rate $\gamma$ from 5~$\mu$eV to 200~$\mu$eV at the
same boundary.  In the outer area, the pump no longer reaches
threshold~(\ref{eq:threshold}), however, the $\Sigma_\pm$ and $\Pi$
states are still unstable.  Consequently, the condensate wave vector
is uncertain and the field has to be inhomogeneous on the scale
$a \sim k_0^{-1}$, where $\hbar^2 k_0^2 / 2m = D + g/2$ and $m$ is the
polariton mass near $k = 0$.  In the one-dimensional case, this would
have resulted in a stiffly ordered dipolar
network~\cite{Gavrilov18-prl}, but the two-dimensional system has more
freedom and arranges itself into a labyrinthine structure like those
discussed in Ref.~\cite{Gavrilov16}.  It has $S_3 \sim \pm 1$ at the
intensity maxima and $S_2 = +1$ at the minima and thus can be thought
of as a host of vortex filaments in which vortices are tightly coupled
(frozen) and do not manifest themselves in a usual manner.  The inner
disk-shaped area contains a single vortex at the center, two curved
$\Xi$ segments with $S_2 \sim -1$, and two straight filaments with
$S_2 \sim +1$ connected to the outer labyrinth.  This pattern rotates
at a constant angular velocity and represents an instance of unsteady
but fully established and self-consistent polariton states.  It is
``driven'' by the instability of the $\Xi$ segments which drift in the
same angular direction.  Such picture resembles Yin and Yang, the
symbol of duality.

In summary, we have found a new class of many-body polariton states
which are spontaneously formed owing to parity breakdown under
resonant excitation.  They are analogous to quantized vortices in
Bose-Einstein condensates and superconductors in that they feature
phase singularity and a BKT-like crossover.  However, their collective
properties are different because of a large-scale instability of the
borderline between opposite $\Omega$ domains.  The rotational symmetry
of each vortex is broken and the ordering scale on which it has a
pronounced effect on its environment appears to be much greater than
the healing length.  In certain cases a dilute Bose gas even displays
an effectively rigid-body type of rotation.  To create such states,
one needs a microcavity in which decay rate $\gamma$ is much smaller
than the spin coupling rate.  The excitation threshold
$f_1$~(\ref{eq:threshold}) is expected to decrease with $\gamma$,
which gives reason to hope that the discussed phenomena are feasible
even in very high-$Q$ microcavities.  Notice, however, that the system
considered is not reduced to an equilibrium Bose-Einstein condensate
in the limit of $\gamma \to 0$ and $f_1 \to 0$ because the $\Omega$
states always oscillate at the pump frequency.

\begin{acknowledgments}
  I am grateful to V.\,D.~Kulakovskii, S.\,G.~Tikhodeev, and
  N.\,A.~Gippius for stimulating discussions.  The work was supported
  by the Russian Science Foundation grant 16-12-10538.
\end{acknowledgments}



%

\end{document}